
\input epsf
\input amstex
\loadbold
\documentstyle{amsppt}
\magnification=\magstep 1
\hsize29pc
\vsize42pc
\baselineskip=24truept
\def\res{\text{res}\,}
\def\Jac{\text{Jac}\,}
\def\tr{\text{tr}\,}
\def\sn{\text{sn}\,}
\def\Q{\Cal Q}
\def\L{\Cal L}
\def\g{\gamma}
\def\Z{\Bbb Z}
\def\H{\bar{H}}
\def\bwp{\bar{\wp}}

\def\te{\tilde{e}}
\def\tL{\tilde{\Cal L}}

\def\[{\left[}
\def\]{\right]}
\def\({\left(}
\def\){\right)}
\topmatter
\title  ON EXPLICIT PARAMETRISATION OF SPECTRAL CURVES
FOR   MOSER-CALOGERO PARTICLES AND ITS APPLICATIONS\endtitle
\author K.L.  Vaninsky\endauthor
\affil Department of Mathematics\\
Kansas State University\\
Manhattan, KS 66502
\endaffil
\email vaninsky@math.ias.edu\endemail
\thanks  The work is  supported by NSF grant DMS-9501002\endthanks
\abstract
The system of $N$ classical particles on the line
with the  Weierstrass $\wp$
function as potential is known to be completely integrable. 
Recently D'Hoker and Phong found a  beautiful  parameterization 
by the polynomial of degree $N$ of the space of Riemann surfaces  
associated with this system. In the trigonometric limit of the 
elliptic potential these  Riemann surfaces degenerate into rational curves. 
The D'Hoker-Phong polynomial in the limit describes  the 
intersection points of the rational curves.  We found an 
explicit determinant representation of the polynomial in the trigonometric  case.  We 
consider applications of this result to the theory of Toeplitz determinants 
and to geometry of the spectral  curves. We also   prove  our  earlier conjecture
on the asymptotic behavior of the ratio 
of two symplectic volumes when the number of particles  tends to infinity.
\endabstract
 
\endtopmatter
\rightheadtext{ON EXPLICIT PARAMETRISATION OF SPECTRAL CURVES}
\document
\subhead 1. Introduction\endsubhead 
Since its discovery in the middle of seventies by Moser and Calogero, the 
system of classical particles on the line with the Hamiltonian 
$$
H=\sum\limits_{n=1}^{N} {p_n^2\over 2} - {m^2 \sigma^2\over 2}
\sum\limits_{n,r=1}^{N} \wp(q_n-q_{r}),\quad\quad \text{where} \;\; 
\sigma^2=\pm 1,\;\;  m>0,
$$ 
has attracted a lot of attention from 
mathematicians and physicists. This system is completely integrable in the 
sense of Liouwille, {\it i.e.}, it has $N$ commuting integrals of motion. 
Algebro-geometrical methods in the study of this system were introduced 
by Krichever, \cite{K1}. He showed that the   Riemann surfaces 
--- spectral curves, associated with this system 
are $N$--sheeted covers of the elliptic curve of the potential, the 
Weierstrass $\wp$ function. The geometry of such covers is fairly complicated and 
was not understood well until now.

In the papers of Seiberg--Witten and Donagi--Witten, 
\cite{SW,DW},
this system was considered in connection with Super-Symmetric Yang-Mills 
quantum field theory. Degeneration of the elliptic potential into 
a trigonometric one is 
interpreted by physicists as a classical limit of the quantum system. In 
such limit genus--$N$ spectral curves degenerate into genus--0 curves.  
Recently D'Hoker and Phong, \cite{DP}, found a beautiful parameterization by a 
polynomial $\Q(p)$ of the space of spectral curves for particles with 
elliptic potential. In the trigonometric limit the zeros $k_n$ of the 
polynomial 
$\Q(p)=m^N\prod^{N}_{n=1}(p-k_n)$ describe intersection points of the limiting 
rational curve.  We have found a determinant representation of the 
polynomial $\Q(p)$ in this limiting trigonometric case. Our description becomes 
completely  explicit when the mechanical system is at the ground state: all 
particles at rest, placed equidistantly on the circle. In this case $\Q(p)$   
can be computed explicitly  using the determinant representation. This provides us 
with  complete information about the spectral curve. In all other cases, such as 
an elliptic or a system not at the ground state, the curve can be viewed as a 
deformation of this simple case. 

In this paper we 
present a simple proof of the D'Hoker-Phong theorem, our determinant  
representation of $\Q(p)$ and some of its' various applications.
The paper is organized as follows. In section 2 we present some well-known 
facts about particles with elliptic potential. In section 3 we give a simple 
proof of the D'Hoker--Phong theorem and prove our main result: an explicit 
formula for the polynomial $\Q(p)$. In the three subsequent sections we 
consider various applications of our result: to the theory of Toeplitz 
determinants (section 4), to the geometry of spectral curves (section 5) and finally 
to the asymptotic behavior of the ratio of two symplectic volumes, when the 
number of particles tends to infinity (section 6).

\subhead 2. Commutator formalism. Riemann surface\endsubhead
The $N$-particle Hamiltonian 
$$
H=\sum\limits_{n=1}^{N} {p_n^2\over 2} - {m^2 \sigma^2\over 2} 
\sum\limits_{n,r=1}^{N} \wp(q_n-q_{r}),
$$
produces the equations of motion
$$
\align
\overset\bullet\to q_n & =\;\; {\partial H\over \partial p_n}=p_n,
\quad\quad\quad\quad\quad\quad\quad\quad \quad\quad
n=1,\hdots, N, \tag 2.1\\
\overset\bullet\to p_n & =-{\partial H\over \partial q_n}=m^2 \sigma^2
\sum\limits_{r\neq n} \wp'(q_n-q_{r}), \quad\quad n=1,\hdots, N.
\endalign
$$
For attractive particles the parameter $\sigma=1$ and for repulsive particles $\sigma=i$; 
the parameter $m>0$ plays the role of mass. 

For $m=2$ the $N$-particle system can be embedded into elliptic solutions of 
the Kadomtzev-Petviashvilli (KP)  equation. The key step is the following theorem

\proclaim{Theorem 1}\cite{K1}. The equations
$$
\align
& \[\sigma \partial_y -\partial_x^2 + 2 \sum\limits_{n=1}^{N} \wp(x-q_n(y))\]
\psi =0, \\
\psi^{+} & \[\sigma \partial_y -\partial_x^2 + 2 \sum\limits_{n=1}^{n}
\wp(x-q_n(y))\] =0
\endalign
$$
have solutions of the form
$$
\align
\psi(x,y,k,z) &=\sum\limits_{n=1}^{N}a_n(y,k,z) \Phi(x-q_n,z)
  e^{kx+ \sigma^{-1}k^2 y},\\
\psi^{+}(x,y,k,z) &=\sum\limits_{n=1}^{N}a^{+}_n(y,k,z)
  \Phi(-x+q_n,z) e^{-kx- \sigma^{-1} k^2 y},
\endalign
$$
where
\footnote"*"{$\sigma(z)$ denotes the Weierstrass function.
See section 7 for definitions.}
$\Phi(q,z)={\sigma(z-q)\over \sigma(z)\sigma(q)}e^{\zeta(z) q}$,
if and only if the functions $q_n(y)$ satisfy the system of equations (2.1).
\endproclaim
The operator $L_2$ with the elliptic potential 
$$
L_2=\partial_x^2-u(x,y)=\partial_x^2- 2 \sum\limits_{n=1}^{N} \wp(x-q_n(y))
$$
enters into the commutator formula
$$
\[\sigma \partial_y-L_2,\partial_t-L_3\]=0, 
$$
with $L_3=\partial_x^3-{3\over 2} u\partial_x -w$.
These imply that $u$
satisfies the KP equation
$$
{3\over 4}\sigma^2 u_{yy}=(u_t+{3\over 2} uu_x -{1\over 4}u_{xxx})_x.
$$

For arbitrary mass $m > 0$,  the system (2.1) is equivalent to the 
matrix equation $\sigma\overset\bullet\to L=\[L,M\]$, 
with $N\times N$ matrices $M$ and $L$ having  entries
$$
\align
L_{nr}& = \sigma p_n \delta_{nr} +m\Phi(q_n-q_r,z)(1-\delta_{nr}),\\
M_{nr}& = \(-\wp(z)+ m\sum\limits_{s\neq n}\wp(q_n-q_s)\)\delta_{nr}
+ m \Phi'(q_n-q_r,z) (1-\delta_{nr}). \endalign
$$
The matrix $L$ can be simplified  using a gauge transformation
$$ L=G\tilde L G^{-1},$$
where $G_{nr}=e^{\zeta(z)q_n}\delta_{nr}$ and
$\tilde L_{nr}=\sigma p_{nr}\delta_{nr}+m\Phi_0(q_n-q_m,z)(1-\delta_{nm})$,
$\Phi_0(q,z)=\frac{\sigma(z-q)}{\sigma(z)\sigma(q)}$.
The spectrum of $L$ is preserved and this determines the curve 
$$
\Gamma\equiv\{(k,z):  R(k,z)=\det (L +mk) =\det  (\tilde L+m k)= 0\}.
$$
We denote by lower and upper Greek letters  
the points $(k,z)$ on the curve $\Gamma$, for example, $\gamma,\Pi$, {\it etc.}.

{\bf Remark.} An exchange of positions and velocities of particles
labeled, say, $n$ and $r$ does not affect the curve.
It simply permutes the
$n$-th and $r$-th columns and rows of the matrix $L$.

The following facts about the curve can be found in \cite{K1}.

\noindent
{\it (i.)} The determinant  $R(k,z)$ can be written in the form
$$
R(k,z)=\sum\limits_{n=0}^{N} r_n(z) k^n,
$$
where $r_n(z)$ are elliptic functions of $z$. This  means that
$\Gamma$
is an $N$--sheeted covering of the elliptic curve $\Gamma_0$.

\noindent
{\it (ii.)} In the vicinity of zero  
$$
R(k,z)= m^N\[k-\({N-1\over z}+k_1^{(0)} + \hdots \)\]
\prod\limits_{n=2}^{N} \[k-\(-{1\over z} +k_{n}^{(0)} +\hdots\)\].
$$
The points $\Pi_{n},\;\;n=1,\hdots,N$ above $z=0$ are called infinities. 
The infinity $\Pi_1$ corresponds to the ``upper sheet'', where $k(z) 
= {N-1\over z}+ O(1)$.

\noindent
{\it (iii.)} In the elliptic case, for the generic configuration of $N$
particles the genus of the
curve $\Gamma$ is  $N$.  In the trigonometric  case the  
curve is  rational.

The following lemma describes symmetries of the matrix  $L$. We assume that $2\omega$ is 
real and $2\omega'$ is pure imaginary. 

\proclaim{Lemma 2}  $(i).\quad \sigma= 1$. Let
$\tau_1$ is defined as: $\tau_1(k,z)=(\bar{k}, \bar{z}).$
Then $$\(L+mk\)(\tau_1\g)=\overline{\(L+mk\)(\g)}.$$ 
$(ii).\quad \sigma= i$. Let $ \tau_i$ is defined as: 
$ \tau_i(k,z)=(-\bar{k},-\bar{z}).$
Then $$\(L+mk\)(\tau_i\g)=-\overline{\(L+mk\)^{T}(\g)}.$$
\endproclaim
\demo\nofrills{Proof.\usualspace} (i). The statment follows from the  identities  
$$
\align
(L+mk)_{nn}(\tau_1\g)&=p_n+ m \overline{k}(\g) =
\overline{p_n+mk(\g)}= \overline{(L+mk)_{nn}(\g)},\\
(L+mk)_{nr}(\tau_1\g)& =m\Phi(q_n-q_r,\overline{z})= 
\overline{m\Phi(q_n-q_r,z)}=\overline{(L+mk)_{nr}(\g)}.
\endalign
$$

(ii).  The proof is similar to that in the attractive case. 
\qed
\enddemo

\noindent
Lemma 2 implies that in the attractive/repulsive case there exists 
an antiholomorphic involution $\tau_1/\tau_i$ on the curve $\Gamma$. 

For some special configurations of particles there is also an 
additional symmetry.

\proclaim{Lemma 3} Let $\tau_-$ is defined as:  $\tau_-(k,z)=(-k,-z)$. 
If $p_n=0$
for all $n=1,\hdots, N$; then 
$$\(L+mk\)(\tau_-\g)=-\(L+mk\)^{T}(\g).$$
\endproclaim
\demo\nofrills{Proof.\usualspace} It is enough to note that
$$
\align
(L+mk)_{nn}(\tau_-\g)& = -mk(\g)= -(L+mk)_{nn}(\g),\\
(L+mk)_{nr}(\tau_-\g)&=m\Phi(q_n-q_r,-z)=-m\Phi(q_r-q_n,z)=-(L+mk)_{rn}(\g).
\endalign
$$
\qed
\enddemo

\noindent
Lemma 3 implies that if $p\equiv 0$, then there is an involution $\tau_-$ on 
the curve. 

\subhead 3. D'Hoker-Phong parametrisation. Determinant Formula \endsubhead
Following D'Hoker-Phong, \cite{DP}, let us introduce
$$
h_k(z)={\partial_z^k \theta_1({z\over 2\omega})\over
\theta_1({z\over 2\omega})},
\quad\quad\quad k=0,1,\hdots.
$$
The function $h_1(z)=\zeta(z)-\frac{\eta}{\omega}z$ has periodicity
properties
$$ 
\align
h_1(z+2\omega)& =h_1(z), \\
h_1(z+2\omega')& =h_1(z)-\frac{i \pi}{\omega}. 
\endalign
$$
Let us also introduce a multivalued function
$p(\g)$ on the curve $\Gamma$ as
$$ p(\g)\equiv k(\g)+h_1(z(\g)). $$
The function $p(\g)$ has a simple pole at $\Pi_1$  only and does not
change under the shift $z\to z+2\omega$.
It is defined up to an integer multiple of $\frac{i\pi}{\omega}$
which corresponds to the increment of $h_1$ under the vertical shift
$z\to z+2\omega'$.  It is called {\it quasimomentum}, because
$\psi(x,y,\g)$ defined in Theorem 1 has the Bloch property in the $x$-variable:  
$$ \psi(x+2\omega,y,\g)=e^{p(\g)2\omega} \psi(x,y,\g). $$
So far we represented   $\Gamma$ as a curve 
in the direct product of the $k$ and $z$ variables. Now we want to write an equation for  
$\Gamma$ in the direct   product of $p$ and $z$.

\proclaim{Theorem 4} \cite{DP}. There exists  a polynomial $\Q(p)$ of degree 
$N$, 
$$
\Q(p)=m^N\sum\limits_{n=0}^{N}p^{N-n}Q_{n}=
m^N\prod^N_{n=1}(p-k_n),
$$
such that
$$
{\theta_1\({z-\partial \over 2\omega}\)\over
   \theta_1\({z\over 2\omega}\)} \Q(p)
= \sum\limits_{n=0}^{\infty} {h_n(z) \over n!} (- \partial)^n \Q(p)
=R(p-h_1(z),z).
$$
\endproclaim
\demo\nofrills{Proof.\usualspace}\footnote"*"{This short proof was communicated 
to me by I. Krichever}
Performing the  change of variables $k(\g)=p(\g)- h_1(z(\g))$ we have  $R(p-h_1(z),z)=
m^N\prod^N_{n=1}\(p-p_n(z)\)$.
The roots $p_n(z)$ do not change under the shift $z\to z+2\omega$.
They transform as
$$p_{s_r}(z+2\omega')=p_r(z) -\frac{i\pi}{\omega} $$
under the vertical shift $z\to z+2\omega'$;
$(s_1,\dots,s_N)$ is some permutation  of sheets.
Similarly the operator 
$$\frac{\theta_1\(\frac{z-\partial}{2\omega}\)}{\theta_1(\frac{z}{2\omega})}
$$
does not change under the horizontal shift 
and for the vertical shifts we have
$$ \frac{\theta_1\(\frac{z+2\omega'-\partial}{2\omega}\)}
{\theta_1\(\frac{z+2\omega'}{2\omega}\)}
= \frac{\theta_1\(\frac{z-\partial}{2\omega}\)}{\theta_1
  \(\frac{z}{2\omega}\)} e^{\frac{\pi i}{\omega}\partial}\ .
$$
Because of these, the polynomial
$$ 
\Q(p,z)\equiv m^N \sum^N_{n=0} p^{N- n} Q_n (z)= 
\frac{\theta_1\(\frac{z}{2\omega}\)}{\theta_1\(\frac{z-\partial}{2\omega}\)}
R(p-h_1(z),z)
$$
has elliptic coefficients $Q_n(z)$. 
The coefficients  of the polynomial on the right hand side 
$$ R\(p-h_1(z),z\)
= m^N\[ p^N+ \( \sum^N_{n=1}- p_n(z) \)p^{N-1  }+\dots
+ \(\prod^N_{n=1}-p_n (z)\) \]
$$
have singularities at $z=0$ not higher than a simple pole. 
The operator \newline $\theta\(\frac{z-\partial}{2\omega}\)/
\theta\(\frac{z}{2\omega}\)$ does not decrease the  order of
a singularity. Therefore, the elliptic coefficients $Q_n(z)$   have 
a singularity at $z=0$ not higher  than a simple pole, and these singularities 
do not depend on $z$ at all.
\qed
\enddemo

When we want to emphasize the dependence of all functions on
modular parameter $\tau$ we write $h(z)=h(z\vert\tau)$,
{\it etc.}
In the trigonometric limit $\tau\to +i \infty$ an
elliptic curve $\Gamma_0$ degenerates into a  sphere
with two punctures, parametized by
$w=e^{-\frac{i\pi z}{\omega}}$. North ($\beta$) and
south ($\nu$) punctures correspond to $w=\infty$ and $w=0$.
The function $p(\g)$  can be used as a global parameter on the limiting rational curve.
The equation of the curve
$$ \theta_1\(\frac{z-\partial}{2\omega_1}\bigg\vert\tau \)
   \Q(p\vert\tau)=0,
   $$
using the definition of $\theta_1$ can be written as
$$ \sum_n\(-1\)^n h^{n(n-1)}
   w^{-n}\Q\(p+\frac{i\pi}{2\omega}- \frac{i\pi n}{\omega}\bigg\vert\tau\)=0.
   $$
In the limit $\tau= +i\infty$ it becomes
$$ \Q\(p+\frac{i\pi}{2\omega}\bigg\vert{+ i\infty}\)
   -w^{-1}\Q\(p-\frac{i\pi}{2\omega}\bigg\vert{+i\infty}\)=0.
   $$
For each point $\nu_n \in\Gamma$, $n=1,\dots,N$
above $w=0$ there exists a point $\beta_{s_n}\in\Gamma$,  above $w=\infty$ 
such that
$$ p\(\nu_n\)-\frac{i\pi}{2\omega}=k_n
   =p\( \beta_{s_n}\)+\frac{i\pi}{2\omega}. \tag{3.1}
   $$
It is easy to check that $\nu_n$ and $\beta_{s_n}$ correspond
to a simple crossing.

Let us introduce the $N\times N$ matrix $\L$ such that
$$ \L_{nr}=\sigma p_n \delta_{nr} +m R(q_n-q_r)(1-\delta_{nr}), \quad \quad \text{where} 
\quad R(q)\equiv \frac{\pi}{2\omega}\frac{1}{\sin\frac{\pi q}{2\omega}}. $$

\proclaim{Theorem 5}
The polynomial $\Q(p\vert\tau)$ for $\tau=+i\infty$ is given by the formula
$$ \Q(p\vert +i\infty) =\det\,\(\L+mp\). $$
\endproclaim

\demo{Proof}
It is enough to prove that
$$ \lim_{\tau\to +i\infty}
   \det\, \( \widetilde L(z\mid\tau)+ mp \) \bigg\vert_{z=\frac{\tau}{2}}=
    \det\, \( \L +mp \),
   \tag3.1 $$
$$ \lim_{\tau\to +i\infty} h_1(z\vert\tau)\bigg\vert_{z=\frac{\tau}{2}}
   =-\frac{i\pi}{2\omega},
   \tag3.2 $$
$$ \lim_{\tau\to +i\infty}
   \frac{\theta_1\(\frac{z-\partial}{2\omega}\vert \tau \)}{\theta_1\(
\frac{z}{2\omega} \vert\tau\)}
   \bigg\vert_{z=\frac{\tau}{2}} =e^{\frac{i\pi}{2\omega}\partial}.
   \tag3.3 $$
After that, the D'Hoker-Phong formula taken at $z=\frac{\tau}{2}$:
$$ \frac{\theta_1\(\frac{z-\partial}{2\omega}\vert \tau \)}
     {\theta_1\(\frac{z}{2\omega}\vert \tau \)}\bigg\vert_{z=\frac{\tau}{2}}
     \Q(p\vert \tau)  
     = \det\(\widetilde L(z\vert \tau )+m (p-h_1(z)\)
     \bigg\vert_{z=\frac{\tau}{2}}
     $$
implies the statement, when $\tau \rightarrow + i\infty$. 

In order to prove (3.1), note that
$$ \Phi_0(q,z\vert \tau)\vert_{z=\frac{\tau}{2}}
   =\frac{1}{\sn (q\vert\tau)}e^{-\eta'q}
   $$
and
$$ \lim_{\tau\to +i\infty} \frac{1}{\sn(q\vert\tau)} =R(q). $$

To prove (3.2), note that
$$ \partial^k_z \theta_1 \(\frac{z}{2\omega}\)
   = i\sum_n (-1)^n h^{\(\frac{2n-1}{2}\)^2}
   \[\frac{i\pi}{2\omega}(2n-1)\]^k
   e^{i\pi \frac{z}{2\omega}(2n-1)}.
   $$
Therefore,
$$ \partial^k_z \theta_1 \(\frac{z}{2\omega}\bigg\vert\tau\)
   \bigg\vert_{z=\frac{\tau}{2}}
   =i\[-\frac{i\pi}{2\omega}\]^k e^{-\frac{i\pi\tau}{4}}
   + O \( e^{\frac{i3\pi \tau}{4}} \),
   $$
when $\tau \longrightarrow +i \infty$. Finally, 
$$ \lim_{\tau\to +i\infty} h_k(z\vert\tau)\bigg\vert_{z=\frac{\tau}{2}}= 
   \[-\frac{i\pi}{2\omega}\]^k.
   $$
This completes  the proof of (3.2). 

In order to prove the last identity (3.3) we use the previous result 
$$ \lim_{\tau\to +i\infty}
   \frac{\theta_1\(\frac{z-\partial}{2\omega}\vert \tau  \)}
{\theta_1\(\frac{z}{2\omega}\vert \tau \)}
   \bigg\vert_{z=\frac{\tau}{2}}
   =\lim_{\tau\to +i\infty} \sum^\infty_{k=0} \frac{h_k(z\vert \tau)}{k!}(-\partial)^k
   =\sum^\infty_{k=0} \frac{1}{k!}
   \(\frac{i\pi}{2\omega}\partial\)^k
   =e^{\frac{i\pi}{2\omega}\partial}.
   $$

In fact, only the first  $N+1$ terms  in all sums matter,  since the  degree of each 
polynomial in $p$ does not exceed $N$. This justifies interchange of the 
limit and sum. The last identity (3.3) is proved.
\qed
\enddemo

There are two essentially distinct cases. For $\sigma=i$ the matrix $\L$ is 
skew-adjoint and all $k_n$, roots of the polynomial $\Q(p)$,  are pure imaginary. 
For $\sigma=1$ the 
skew-adjointness is lost, but similar to Lemma 3, if $p\equiv 0$ the 
skew-adjointness of $\L$ is restored and all $k_n$ are pure imaginary.

\subhead 4. Spectrum of the Toeplitz operator\endsubhead 
For a finite number of repulsive particles ($\tau \leq +i\infty$) there 
exists a point  or set of points in the phase space where the Hamiltonian 
$H(q,p)$, bounded from below,  
achieves its minimum. Such a point $\overline{p}_n, \overline{q}_n, \; n=1,\hdots, N$ is 
called a {\it ``ground state''}. In our case the Hamiltonian is rotationaly 
invariant and ground state is a one-parameter family:  $\overline{p}_n=0,\; 
\overline{q}_n-\overline{q}_{n-1}=\frac{2\omega}{N}$, for all $n=1,\hdots, N$.

Obviously, the matrix $\L$ at the ground state is Toeplitz and skew--symmetric.
Let us intoduce $e^T=(e^1,\hdots,e^N)^T,$ an eigenvector of  $\L:\;  \(\L+km\)e=0$ 
corresponding to the eigenvalue $k$ and normalized such that $e^1=1$. 
\proclaim{Lemma 6} Let $\sigma=i$. If the  system is at the ground state, then the 
eigenvalues and eigenvectors  of the matrix $\L$ are  
given by the formulas
$$
k_n =\frac{\pi}{2\omega i} (N+1-2n),\quad \quad \quad n=1,\hdots,N;
$$
and
$$
e_{n}^{j} = e^{i\frac{\pi}{N}(j-1)(1-2n)},\quad\quad j=1,\hdots,N;\quad 
n=1,\hdots, N.
$$
\endproclaim

To prove the Lemma we introduce some formulas from  Fourier analysis on the 
additive group $G_N=\Z/N\Z$ of the ring of residues modulo $N$. 

\noindent
{\it i.} Characters $\mu\in \hat G_N$ are all defined as $\mu_{n}=e^{i\beta_{n}},\,
\beta_{n}=\frac{2\pi}{N}(n-1),\; n=1,\hdots,N.$ The pairing is: for any $s\in G_N,\; 
\mu \in \hat G_N:\; 
(s,\mu)=\mu^s$. 

\noindent
{\it ii.} For any complex function $a=\{a_s,\, s\in G_N\}$ its Fourier transform 
is $\hat a(\mu)= 
\sum\limits_{s\in G_N} a_s (s,\mu),$ and the inverse transform is 
$a_s=\frac{1}{N}\sum\limits_{\mu\in \hat G_N} \hat a(\mu) (-s,\mu).$ 

\noindent
{\it iii.} For any complex functions $a,b$ the Plancherel identity holds:
$
\sum\limits_{s\in G_N} a_s \overline{b}_s= \frac{1}{N} \sum\limits_{\mu\in \hat G_N} 
\hat a(\mu) \overline{\hat b(\mu)}.
$
\demo{Proof} To simplify calculations let $m=\frac{2\omega i}{\pi}$ and also re-scale the 
spectral parameter 
$k\longrightarrow \frac{\pi}{2\omega i}k$. General case easily 
follows from this.

{\it Step 1.} Let $e_+^T=(e^1,\hdots,e^N)^T$ is an eigenvector 
corresponding to the eigenvalue $k$. We will prove that $e_-^T=(e^N,\hdots, e^1)^T$ is 
an eigenvector corresponding to the eigenvalue $-k$. Indeed, for any $j$
$$
\sum\Sb r\neq j\\ r\neq N-j+1 \endSb \L_{jr}e^r_++ \L_{j,N-j+1}e_+^{N-j+1}+ke_+^j=0.
$$
Then, 
$$
\sum\Sb r\neq j\\ r\neq N-j+1 \endSb \L_{j,N-r+1}e^{N-r+1}_++ \L_{j,N-j+1}e_+^{N-j+1}+
k e_+^j=0.
$$
At the ground state $\L_{N-j+1,N-r+1}=-\L_{jr}$. Introducing $e_-^n=e_+^{N-n+1}$, 
we obtain
$$
-\sum\Sb r\neq j\\ r\neq N-j+1 \endSb \L_{N-j+1,r}e^r_-- \L_{N-j+1,j}e_-^{j}-
(-k)e_-^{N-j+1}=0.
$$ 
The statement is proved.

{\it Step 2.} For $j \neq r$ we have 
$$
\L_{jr}=mR(\overline{q}_j-\overline{q}_r)=\frac{(\sqrt{2}i)^2 \mu^{1/2}_j \mu^{1/2}_r}
{\mu_j-\mu_r}.
$$
Let $D,\tL$ be the  $N\times N$ matrices with entries
$$
\align
D_{jr}& =\sqrt{2}i \mu_j^{1/2}\delta_{jr},\\
\tL_{jr}& =-\frac{k}{2}\mu_j^{-1}\delta_{jr}+\frac{1}{\mu_j-\mu_r}(1-\delta_{jr}).
\endalign
$$
Then, $\L+mk=D\tL D$ and the eigenvalue problem can be written as
$$
(\L+mk)e=D\tL D e= \tL \te=0,
$$
where $\te=De$. For any $j$
$$
-\frac{k}{2}\mu_j^{-1}\te^j +\sum\limits_{r\neq j}\frac{\te^r}{\mu_j-\mu_r}=0.
$$
Introducing $u_j=\mu_j^{-1}\te^j$ and using $\mu_j/\mu_r=\mu_{j-r+1}$, we obtain 
$$
-\frac{k}{2}u_j +\sum\limits_{r\neq j} \frac{u_r}{\mu_{j-r+1}-1}=0.
$$
This is a convolution equation which can be solved using Fourier analysis. Introducing
$$
\hat U(\mu)=\sum\limits_{s\in G_N} u_s(s,\mu),\quad\quad\quad
\hat M(\mu)=\sum\limits_{s\in G_N-\{N\}} \frac{1}{\mu_{s+1}-1}(s,\mu),
$$  
we arrive at the main equation
$$
\hat U(\mu)(2\hat M(\mu)-k)=0.
$$

{\it Step 3.} Note $\mu_{n}=\mu_2^{n-1}$. For $n \geq 2$ we have 
$$
\align
\hat M(\mu_n) &= \sum\limits_{s\in G_N\backslash \{N\}} \frac{\mu_n^s}{\mu_{s+1}-1}\\
&=\sum\limits_{s\in G_N\backslash \{N\}} \frac{\mu_2^{(n-1)s}-1+1}{\mu_2^s-1}\\
&=\hat M(\mu_1) +\sum\limits_{s\in G_N\backslash \{N\}}\sum\limits_{k=0}^{n-2}(\mu_2^s)^k\\
&= \hat M(\mu_1) +  \sum\limits_{k=1}^{n-2}\sum\limits_{s\in G_N\backslash \{N\}} 
(\mu_2^k)^s + \sum\limits_{s\in G_N\backslash \{N\}} 1^s.
\endalign
$$
If $r^N=1$ and $r\neq 1$, then $ r +r^2 +\hdots r^{N-1}=-1$.  Therefore 
$$
\hat M(\mu_n)= \hat M(\mu_1)  + N+ 1 -n.
$$
From this formula we see that the function $\hat M(\mu)$ takes different values at 
the different points of the character group 
$\hat G_N$. Let $k_n=2\hat M(\mu_n),\quad n=1,\hdots, N$ in the 
main equation of  Step 2. Then, $ k_n=k_1 +2(N+1-n)$, for $n \geq 2$. 
The eigenvalue $k_1$ is the smallest and $k_2$ is the largest. Using  Step 1: $-k_1=k_2$,  
we finally obtain
$$
\align
k_1&=-N+1,\\
k_n& =N+3-2n, \quad\quad\quad n=2,\hdots, N.
\endalign
$$

{\it Step 4.}  Let, for any  $n$
$$
\hat U_n(\mu_r)=N\sqrt{2}i e^{i\frac{2\pi}{N}(n-1)}\delta_{nr}, 
\quad\quad \mu_{r}\in \hat G_N.
$$
Inverting the Fourier transform
$$
u^j_n=\frac{1}{N}\sum\limits_{\mu\in \hat G_N}\hat U_n (\mu) (-j,\mu) =
\sqrt{2}ie^{i\frac{2\pi}{N}(n-1)}\mu_n^{-j}.
$$
Using the identities $u^j=\mu_j^{-1}\te^j,\quad \te^j=\sqrt{2}i \mu_j^{1/2}e^j$, we 
arrive at  the formula 
$e^j_n=e^{\frac{\pi}{N}(j-1)(3-2n)}$. Shifting $n\longrightarrow n+1$, 
completes the proof.
\qed
\enddemo

When the number of particles with $m=1$ tends to infinity for the  ground state with 
$v=\frac{2\omega}{N}=1$ the matrix $\L$ has the  limit 
$$
\L_\infty=\left|\matrix 
\format\c&\quad\r&\quad\r&\quad\r&\quad\r\\
0 & -\frac{1}{2}  & -\frac{1}{3} & -\frac{1}{4} & \hdots \\
\frac{1}{2} & 0 & -\frac{1}{2}  & -\frac{1}{3} & \hdots \\
\frac{1}{3} & \frac{1}{2}  & 0  & -\frac{1}{2} & \hdots \\
\frac{1}{4} &\frac{1}{3} & \frac{1}{2}  & 0    & \hdots \\
\vdots& \vdots & \vdots  & \vdots  & \ddots 
\endmatrix \right|
$$ 
The corresponding Toeplitz operator $T_{\varphi}$ has the symbol 
$\varphi(t)=i\pi-it,\quad 0\leq t < 2 \pi$. To emphasize 
the dependence of $k_n$ on $N$ we  write $k_{n,N}$. For 
any continuous function $f$ on $iR^1$ we have
$$
\lim_{N\longrightarrow \infty} \frac{1}{N} \sum\limits_{n=1}^{N} 
f(k_{n,N})=\frac{1}{2\pi} \int\limits_{0}^{2\pi} f(\varphi(t))dt.
$$
Such distribution of eigenvalues is called {\it canonical}, see \cite{W}. 
It is usually proved with the aid of Szeg\"o theorem. Our case is not 
covered by the conditions of the classical theorem, but the  conclusion for this symbol is 
true also.

\subhead 5. Real ovals of anti-involution. Images of infinities\endsubhead
In the trigonometric case the curve $\Gamma$ is rational and 
$p(Q), \; Q\in \Gamma$ can be used as a uniformization parameter. 
On the curve $\Gamma$ anti-involution $\tau_1/\tau_i$ can be 
expressed in terms of the parameter $p$. 
\proclaim{Lemma 7} 
(i).  $\sigma=1$.  Then   $p(\tau_1Q)=\overline{p(Q)}$.  \newline
(ii).  $\sigma=i$.   Then  $p(\tau_i Q)=- \overline{p(Q)}$.

\endproclaim

\demo{Proof} (i). The statement follows from identity 
$$p(\tau_1Q)=\overline{k(Q)}+h_1(\overline{z(Q)}) = \overline{p(Q)}.$$

(ii). Similarly
$$
p(\tau_iQ)=-\overline{k(Q)}+h_1(-\overline{z(Q)})=-\overline{p(Q)}.
$$
\qed
\enddemo
\noindent
In the attractive case, as a corollary of  the Lemma,   we obtain, that $\Im p=0$ is 
a fixed oval of anti-involution $\tau_1$, and in the repulsive case $\Re p=0$ is a 
fixed oval of anti-involution $\tau_i$. 

Consider the repulsive case and let us  look at the images of infinities 
$\Pi$.  For
$n \geq 2$ we have
$$
p(\Pi_n)=k_n(\gamma)+ h_1(\gamma)|_{\gamma=\Pi_n}= -\frac{1}{z} + k_n^{(0)}
+O(z)+ \frac{1}{z}|_{z=0}=k_n^{(0)}.
$$
As is proved in Lemma 5 of \cite{V2},  all $k_n^{(0)}$ are pure
imaginary and therefore all images of infinities $\Pi_2,\hdots,\Pi_N$  lie 
on the fixed oval
$\Re p=0$ of anti-involution $\tau_i$.
For $n=1$ and $\gamma$ in the vicinity of $\Pi_1$
$$
p(\gamma)=k(\gamma)+h_1(\gamma)=\frac{N-1}{z} + O(1) +\frac{1}{z}=
\frac{N}{z}+O(1).
$$
Therefore $\Pi_1$ corresponds infinity of the uniformization parameter $p$. 

For example, for $N=2$ at the ground state  $k_2^{(0)}=0$ and due to Lemma 6 we have 
the fig 1.
\midinsert
\epsfxsize=250pt
\centerline{
\epsfbox{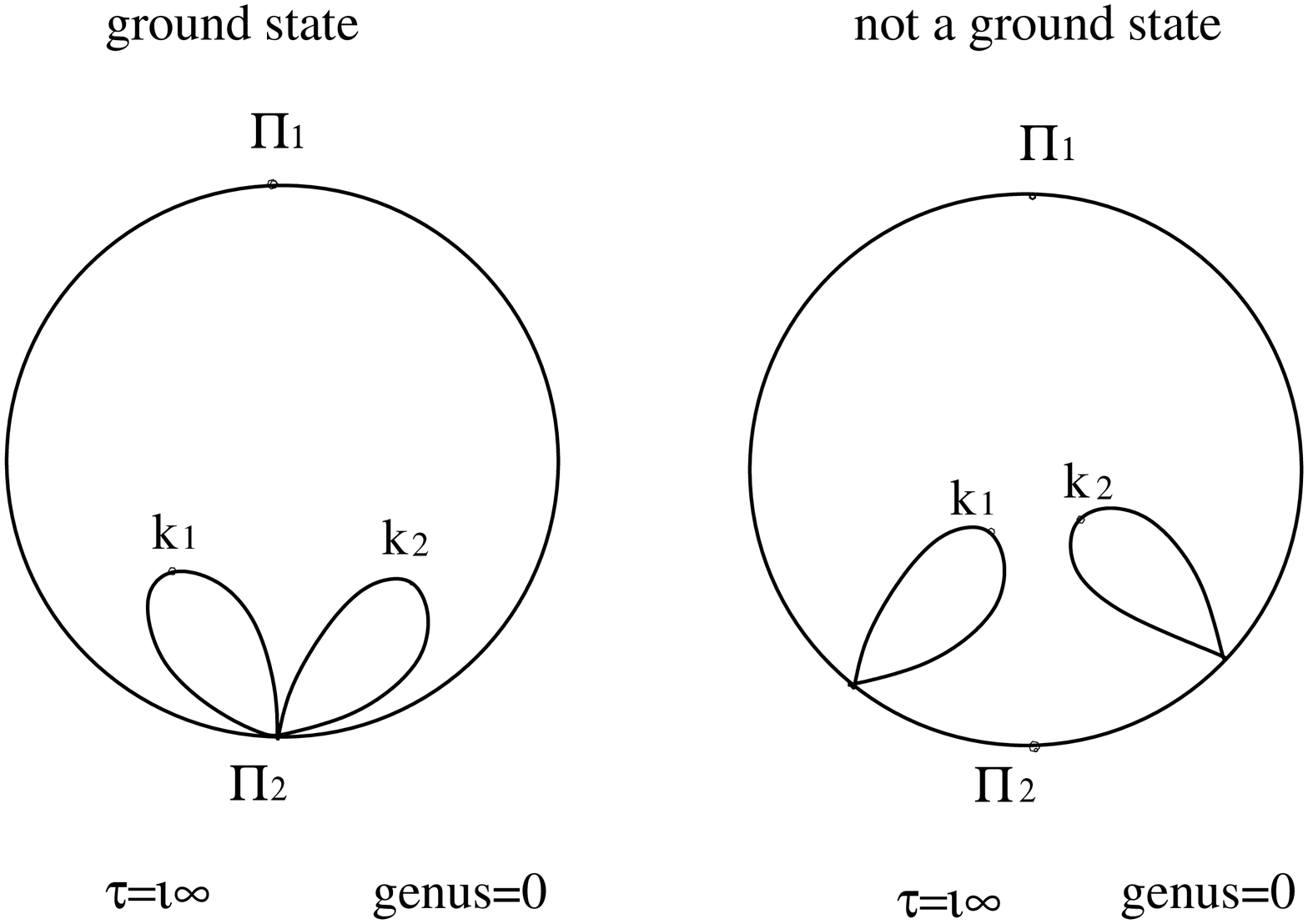}
}
\botcaption{fig. 1 \ \ \ \   \ \ \ \ \ \ \ \ \ \ \ \ \ \ \ \ \ \ \ \ \ \ \ \   fig. 2}
\endcaption
\endinsert
\noindent
When one pumps  energy into the system the loops are perturbed, but $k_n^{(0)}$ 
are still pure imaginary (fig 2). When the  modular parameter $\tau$ changes from 
$+i\infty$ (trigonometric case) to some finite imaginary number (elliptic case), the 
loops ``move inside'' and become fixed ovals around holes, as  is shown in fig 3, 4. 
\midinsert
\epsfxsize=250pt
\centerline{
\epsfbox{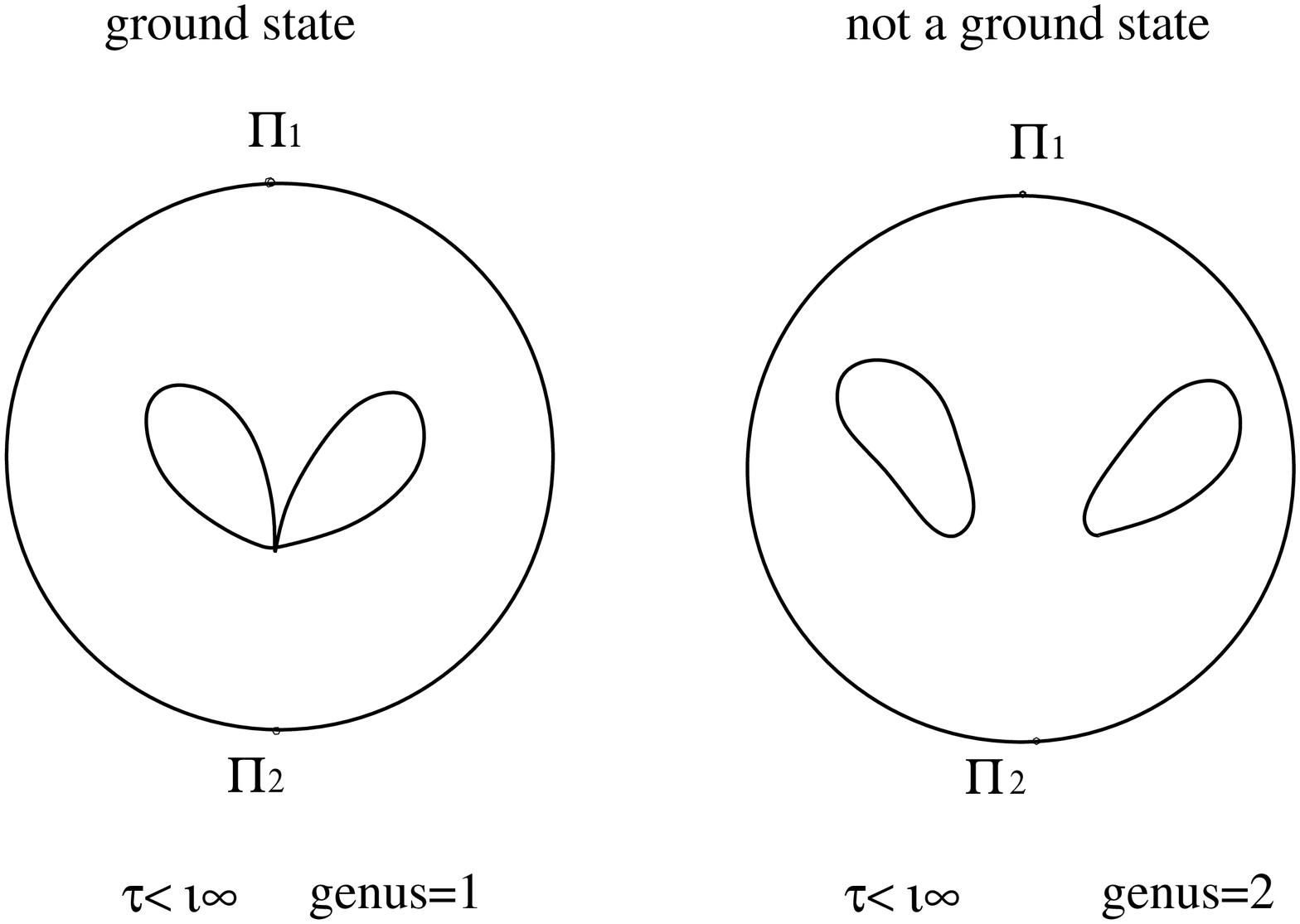}
}
\botcaption{fig. 3 \ \ \ \   \ \ \ \ \ \ \ \ \ \ \ \ \ \ \ \ \ \ \ \ \ \ \ \   fig. 4}
\endcaption
\endinsert
\noindent
For $N>2$ the picture is similar, but $N$ loops are present.

The case of attractive particles is more complicated, 
because skew adjointness of the matrix $\L$ is lost. It is restored  for $p\equiv 0$.

In the trigonometric case, for both repulsive and attractive particles 
the images of infinities $\Pi_2,\hdots,\Pi_N$ 
are {\it moduli} of the spectral curves. We will prove this statement  
in the  elliptic case. First, we derive the trace formula relating total 
momentum $P$ and $k_n^{(0)},\; n=2,\hdots,N$. 

Computing $\sum_{\Pi} \res k^2 \, dz= 0$, we have 
$\sum_{n=1}^{N}k_{n}^{(-1)}k_n^{(0)}=0$. 
Using $k_n^{(-1)}=-1,\; n\geq 2$ and $k_1^{(-1)}=N-1$, we obtain 
$Nk_1^{(0)}=\sum_{n=1}^{N} k_n^{(0)}$. On the other hand  $\tr L= 
\sigma P= -m\sum_{n=1}^{N} k_n^{(0)}$. By comparing the last two expression we 
arrive at  the trace formula
$$
P=\sum\limits_{n=2}^{N}J_n,\quad \quad \text{where}\quad J_n=-k_n^{(0)}
\frac{mN}{\sigma (N-1)}.
$$ 
For another trace formula relating $H$ and $k_n^{(-1)}$ see \cite{V2}. Now, we 
are ready to prove 
\proclaim{Lemma 8} In the elliptic case $J_n,\; n=2,\hdots, N,$ supplemented by  $Q_N$, 
the constant term of the polynomial $\Q(p)$, are moduli of the spectral curves.
\endproclaim
\demo\nofrills{Proof.\usualspace} Using the representation of $R(k,z)$ in the vicinity of 
zero and $h_1(z)=\frac{1}{z} +O(1)$ we have 
$$
R(p-h_1(z),z)=m^N\(p-\frac{N}{z}- k_1^{(0)}+O(z)\)\prod\limits_{n=2}^{N}
\(p-k_n^{(0)}+O(z)\).
$$
Therefore
$$
\underset z=0 \to \res R(p-h_1(z),z)=-Nm^N \prod\limits_{n=2}^{N} (p-k_n^{(0)}).
$$
From the D'Hoker-Phong formula
$$
\sum\limits_{k=0}^{\infty} \underset {z=0} \to \res \frac{h_k(z)}{k!} (-\partial)^k \Q(p)=
-Nm^N  \prod\limits_{n=2}^{N}(p-k_n^{(0)}).
$$
Since $\res h_k(0)=0$ for $k$ even and $\res h_1(0)=1$ we can recover from 
$J$'s  the derivative of $\Q(p)$. The constant term $Q_N$ allows us to  reconstract 
the whole polynomial $\Q(p)$.  
\qed
\enddemo

\subhead 6. Symplectic volumes\endsubhead Let us intoduce in a general 
elliptic case the new Hamiltonian
$$
\H=\sum\limits_{n=1}^{N} {p_n^2\over 2} - {m^2 \sigma^2\over 2}
\sum\limits_{n,r=1}^{N} \bwp(q_n-q_{r}),
$$
where $\bwp=\wp +\frac{\eta}{\omega}$. In the trigonometric  limit $\bwp(q)=
R^2(q)$.

Consider now the trigonometric case. Singularities of the curve $\Gamma$ are simple 
crossings,  as it is shown  in  fig 5.
\topinsert
\topcaption{fig. 5} \endcaption
\epsfxsize=90pt
\centerline{
\epsfbox{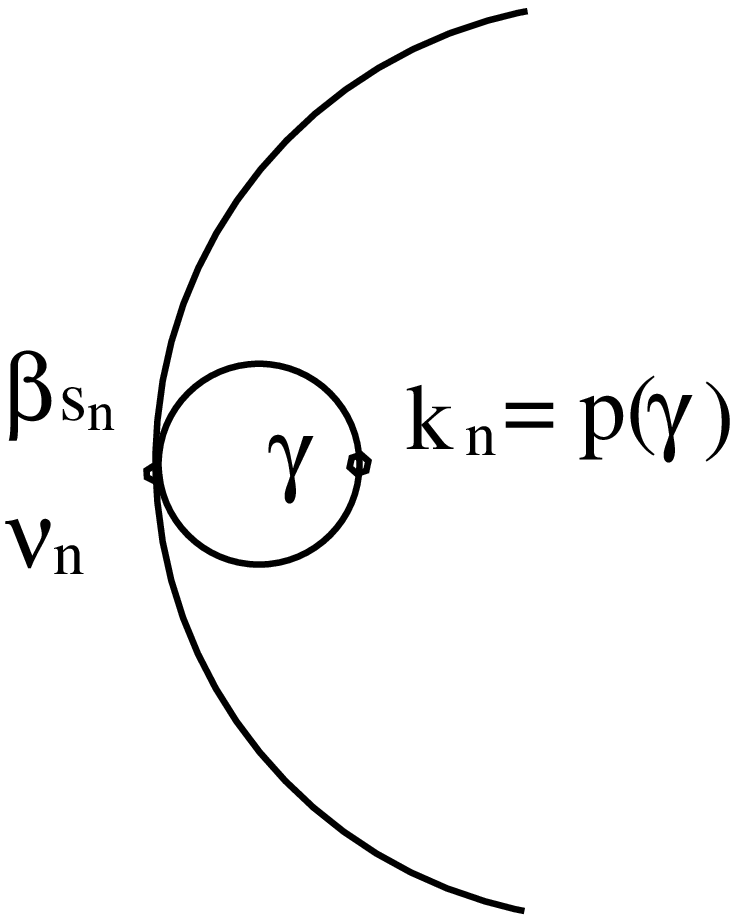}
}
\endinsert
\noindent
\noindent
Form the identity (3.1) is follows that $k(\nu_n)=k(\beta_{s_n})=k_n$. We will 
derive a set of new trace formulas. 

\proclaim{Lemma 9} In the trigonometric case the following formulas hold
$$
\align
P&=\sum\limits_{n=1}^{N}I_n,\quad\quad\quad 
\text{where}\quad  I_n=-\frac{m}{\sigma} k_n,\\
\H&=\sum\limits_{n=1}^{N}I_n',\quad\quad\quad \text{where}\quad I_n'=
\frac{m^2\sigma^2}{2} k_n^2. 
\endalign
$$
\endproclaim
\demo\nofrills{Proof.\usualspace} 
For $\Q(p)$ we have 
$$
\align
\Q(p)&= m^N \prod\limits_{n=1}^{N} (p-k_n)\\
     &=m^N\[p^N +p^{N-1}\(\sum\limits_{n=1}^{N}-k_n\) + 
       p^{N-2}\(\sum_{n,n'=1}^{N} k_nk_{n'}\) +\hdots \].
\endalign
$$
On the other hand  we can write the polynomial $\Q(p)$ as a Fredholm expansion 
for the determinant given by  Theorem 5
$$
\align
\Q(p)&=\det (\L +mp)= (mp)^N +(mp)^{N-1}\(\sum\limits_{n=1}^{N}\sigma p_n\) \\
         +  &(mp)^{N-2}\(\sum_{n,n'=1}^{ N} \sigma^2 p_np_{n'}- 
          m^2R(q_n-q_{n'})R(q_{n'}- q_n)   \) +\hdots .
\endalign
$$
Comparing  coefficients after simple algebra,  we obtain the result.
\qed
\enddemo
 
In the elliptic case a simple crossing  becomes a handle, as  shown in fig 6. 
\topinsert
\topcaption{fig. 6} \endcaption
\epsfxsize=90pt 
\centerline{
\epsfbox{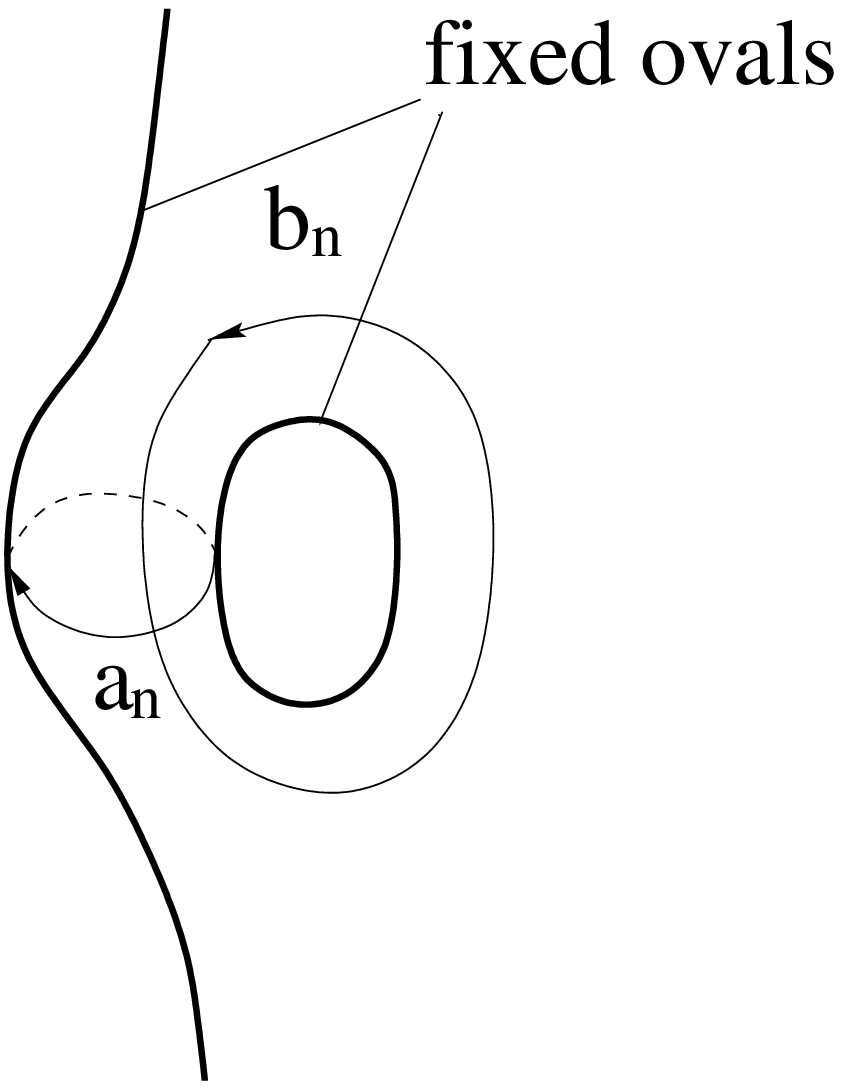}
}
\endinsert
\noindent
In this case one can relate $P,\H$ to the periods of $kdz,k^2dz$ over $a$-circles of $\Gamma$.
\proclaim{Lemma 10} In the elliptic case these formulas hold: 
$$
\align
P&=\sum\limits_{n=1}^{N}I_n,\quad\quad\quad
\text{where}\quad  I_n=-\frac{m}{\sigma} \int\limits_{a_n} k(z) 
\frac{dz}{2\omega},\\
\H&=\sum\limits_{n=1}^{N}I_n',\quad\quad\quad \text{where}\quad I_n'=
\frac{m^2\sigma^2}{2} \int\limits_{a_n}  k^2(z)\frac{dz}{2\omega}.
\endalign
$$
\endproclaim

\demo\nofrills{Proof.\usualspace}\footnote"*"{For another proof see 
\cite{DP}.} Let $c_n$ be  small contours around $\Pi_n,\;n=1,\hdots,N$. 
Using Cauchy' theorem 
$$
\sum\limits_{n=1}^{N}\int\limits_{c_n}k(z)h_1(z) dz + 
\sum\limits_{n=1}^{N}\int\limits_{a_n}k(z)(h_1(z+2\omega') -h_1(z)) dz= 0
$$
Note that $h_1(z+2\omega') -h_1(z)= -\frac{i\pi}{\omega}$ and the second sum 
reduces to
$$
\hdots= \sum\limits_{n=1}^{N}  -\frac{i\pi}{\omega} \int\limits_{a_n} k(z) dz.
$$ 
To evaluate the first  sum we use the asymptotics of $k$ near infinities $\Pi$:
$$
\int\limits_{c_n}k(z) h_1(z) dz =\int\limits_{c_n} \(\frac{k_n^{(-1)}}{z} + 
k_{n}^{(0)} + O(z)\)\times
      \(\frac{1}{z} +O(z)\) dz = 2\pi i k_n^{(0)}.
$$ 
As a result we derive 
$$
\sum\limits_{n=1}^{N} k_n^{(0)}=\sum\limits_{n=1}^{N} \int\limits_{a_n}k(z) 
\frac{dz}{2\omega}.
$$
On another hand  we have 
$$
\tr L =\sigma P=-m\sum\limits_{n=1}^{N}k_n^{(0)}.
$$
The first formula is proved. 

To prove the second trace formula one should integrate $k^2dz$ over the same 
contour.
\qed
\enddemo

The meromorphic 2--forms
$$
\align
\Omega & =-\frac{m}{\sigma}\sum\limits_{n=1}^{N} d k(\g_n)\wedge dz(\g_n) = 
\sum\limits_{n=1}^{N} dI_n\wedge d\phi_n\\
\Omega' & =\frac{m^2 \sigma^2}{2}\sum\limits_{n=1}^{N} d k^2(\g_n)
 \wedge dz(\g_n) = 
\sum\limits_{n=1}^{N} dI_n'\wedge d\phi_n
\endalign
$$ 
on the space of spectral curves can be written following \cite{KP,K2} in a different way, 
namely  
\proclaim{Theorem 11} In the elliptic case the meromophic 2-forms can be 
written as 
$$
\align
\Omega & = -\frac{m}{\sigma} \sum\limits_{n=1}^{N} \res_{\Pi_n} <a^*(\delta L -mk) 
\wedge \delta a> dz,\\
\Omega' & = \frac{m^2\sigma^2}{2} \sum\limits_{n=1}^{N} \res_{\Pi_n} 
k <a^*(\delta L -mk) \wedge \delta a> dz.
\endalign
$$
\endproclaim
The form $\Omega$ can be written as $\sum dq_n\wedge dp_n$, but for the higher form 
$\Omega'$ an explicit expression is not known.

In the paper \cite{V1} we considered a Jacobian 
between symplectic volumes
$$
\Jac \equiv \frac{\overset N\to\wedge \Omega}{\overset N\to\wedge 
\Omega'}=\prod\limits_{n=1}^{N} 
\frac{\partial I_n}{\partial I_N'}.
$$
The quantity $N^{-1}\log \Jac$, when $N\rightarrow \infty$, has a meaning of 
{\it entropy} or,  in probabilistic terms,  {\it rate function}, 
see \cite{L,Va}. 
\proclaim{Lemma 12} For repulsive particles in the trigonometric case 
at the ground state the following asymptotics  hold\footnote"*"{This formula was 
first conjectured  in \cite{V1}.}

$$
\frac{1}{N} \log \Jac = \log N + \log v + \log\frac{2}{\pi m} + o(1), \quad
\text{when}\;\; N\rightarrow\infty;
$$
where $v=\frac{2\omega}{N}$.
\endproclaim
\demo\nofrills{Proof.\usualspace}
Consider the case when $N$ is divisible by 4. At the ground state Lemma 6 
implies
$$
I_n=\frac{m\pi}{Nv}(N+1-2n),\quad\quad\quad I_n'=\frac{m^2 \pi^2}{2N^2 v^2}(N+1-2n)^2.
$$
Therefore,
$$
\Jac= \prod\limits_{n=1}^{N}\frac{\partial I_n}{\partial I_n'} =
\(\frac{2v}{m\pi}\)^N\prod\limits_{n=1}^{N}\frac{N}{N+1-2n}
$$
and
$$
\frac{1}{N}\log \Jac =\log v + \log\frac{2}{m\pi}+  \log N - \frac{2}{N} 
\log(N-1)!!
$$
Using the  formula 
$$
\frac{2}{N} \log(N-1)!! = \log N - 1 +o(1)
$$
we have
$$
\frac{1}{N}\log \Jac =\log v + \log\frac{2}{m\pi}  +1 + o(1).
$$
The remark of section 2 implies that in fact we consider action-angle variables 
on the fundamental domain $q_1< q_2<\hdots < q_N$ of the configuration 
space $[0, 2\omega)^N$. Therefore the  Jacobian has to be multiplied by $N!$. 
This and 
$N^{-1}\log N!= \log N -1 +o(1)$ imply the statement. 

The case of general $N$ can be considered similarly.
\qed
\enddemo

\subhead 7. Appendix\endsubhead
The Weierstrass function $\sigma(z)$ has periods $2\omega$ and $2\omega'$ and is defined as
$$
\sigma(z)=z\prod
 \{(1-{z\over  \boldsymbol\omega}) \exp({z\over \boldsymbol\omega }+{z^2\over 2
\boldsymbol\omega^2}) \},
$$
where $\boldsymbol\omega=2\omega n + 2\omega' n'$ and
$\prod$ is taken with $n,n' \in \Z^1; \;
n^2+n^{\prime 2} >0$; $\sigma(z)$ has zeros at the points of the lattice 
$2\omega n+ 2\omega'n'$. In the vicinity of zero it has the expansion
$$
\sigma(z)=z+ O(z^5).
$$
Under the shift $z\rightarrow z+2\omega$ the function $\sigma(z)$ transforms as 
$$
\sigma(z+2\omega) =-\sigma(z)\exp(2\eta(z+\omega)).
$$
For the shift $z \rightarrow z +2\omega'$ one should replace $2\eta$ by $2\eta'$ in the
formula above.
The functions $\zeta(z)$ and $\wp(z)$ are defined as:
$$
\zeta(z)={d\over d\, z} \log \sigma(z), \quad \quad \wp(z)= - {d\over d\, z} \zeta(z).
$$
They have respective expansions at zero of the form 
$$
\zeta(z) =\frac{1}{z}+O(z^3), \quad \quad 
\wp(z) =\frac{1}{z^2} + O(z^2);
$$
and  periodicity  properties
$$
\wp(z+2\omega) =\wp(z),\quad\quad\quad 
\zeta(z+2\omega)=\zeta(z) +2\eta.
$$
For the shift $z \rightarrow z +2\omega'$ one should replace $2\eta$ by $2\eta'$ in the 
formula above. The  Legendre's identity is  $\eta\omega'-\eta'\omega=\frac{i\pi}{2}$. 

The Jacobi functions $\theta_0, \theta_1$ are defined as 
$$
\align
\theta_0\(\frac{z}{2\omega}\)& =\sum\limits_{n=-\infty}^{+\infty} (-1)^n h^{n^2} 
e^{\frac{i \pi 2n z}{2\omega}},\\
\theta_1\(\frac{z}{2\omega}\)& =\sum\limits_{n=-\infty}^{+\infty} (-1)^n 
h^{(\frac{2n-1}{2})^2} e^{\frac{i\pi (2n-1)z}{2\omega}}, 
\endalign
$$
where $h=e^{i\pi \tau}$. 
The relation between $\theta_1(\frac{z}{2\omega})$ and $\sigma(z)$ is: 
$$
\sigma(z)=\frac{2\omega}{\theta'_1(0)}\theta_1\(\frac{z}{2\omega}\)
e^{\frac{\eta z^2}{2\omega}}.
$$
The Jacobi $\sn(z)$ is defined as 
$$
\sn (z)= 2\omega\frac{\theta_0(0)}{\theta_1'(0)}\frac{\theta_1(\frac{z}{2\omega})}
{\theta_0(\frac{z}{2\omega})}.
$$
For more information see \cite{HC}.

{\bf Acknowledgments.} I would like to thank I. Krichever, V. Korepin, 
H. McKean, H. Widom for  stimulating discussions. My specials thanks to R. Burckel for 
reading the manuscript and suggesting improvements.

\bye